\documentclass{evn2002}
\usepackage{graphicx}
\begin{document}
   \title{The Equatorial Outflow of SS433}

   \author{Z. Paragi\inst{1,2}, I. Fejes\inst{2}, R.C. Vermeulen\inst{3}, 
	   R.T. Schilizzi\inst{1,4}, R.E. Spencer\inst{5}
           \and A.M. Stirling\inst{6}
          }

   \institute{Joint Institute for VLBI in Europe, Postbus 2, 7990~AA Dwingeloo, The Netherlands
	     \and
	     F\"OMI Satellite Geodetic Observatory, P.O. Box 546, H-1373 Budapest, Hungary
	     \and
	     Astron, Postbus 2, 7990~AA Dwingeloo, The Netherlands
             \and
	     Leiden Observatory, Post box 9513, 2300~RA Leiden, The Netherlands
	     \and
             Nuffield Radio Astronomy Laboratory, Jodrell Bank, Macclesfield, Cheshire, SK11 9DL, United Kingdom
	     \and
             CFA, University of Central Lancashire, Preston, PR1 2HE, United Kingdom
             }
   
   \abstract{We present VLBI imaging results for SS433 from the period 1995--2000. An equatorial
     emission region is detected and confirmed in 7 experiments at 1.6 and 5 GHz. The structure
     of the region changes with time on scales of weeks to months. The nature of emission is unknown,
     but it is certainly non-thermal, maybe optically thin synchrotron. The spectrum however
     might be attenuated by a thermal population of electrons. There were previous indications
     for an equatorial outflow from the system. We do detect outward motion of a radio component,
     and estimate a speed of 1200/sin($i$) km/s. Theoretical work and IR observations
     suggest that the mass-loss rate in SS433 is much higher in this region than in the jets. 
     We suggest that the high-mass microquasars might all have equatorial outflows; in fact, it 
     is the dominant form of mass-loss in these systems.
   }

   \authorrunning{Z. Paragi et al.}
   \titlerunning{The Equatorial Outflow of SS433}
   \maketitle

%________________________________________________________________

\section{Introduction}

Microquasars are Galactic X-ray binary systems harbouring a compact 
object (neutron star or black hole). They produce well collimated,
relativistic jets in which energetic electrons are radiating via the 
synchrotron process. SS433 is the brightest 
permanent radio source of this class. Its precessing jets were first 
identified from the Doppler-shifted optical ``moving" lines 
(Margon \& Anderson \cite{M&A89}), and subsequently imaged by MERLIN, 
VLA, and the EVN. In the 90s VLBI arrays became more sensitive, and 
their resolution and imaging fidelity improved significantly. 
This enabled us to study the SS433 radio beams at a level of 
unprecedented details.

The results presented here were observed in nine experiments, listed 
in Table~\ref{Obs}.
The paper briefly summarizes this work, focusing on the equatorial
emission region, that is related to an outflow perpendicular to the
well studied jets (Paragi et al. \cite{NewAR98},\cite{AA99}).

%__________________________________________________ One column table
   \begin{table}
      \caption[]{List of VLBI experiments}
         \label{Obs}
         \begin{tabular}{rlr}
            \hline
            \noalign{\smallskip}
            Date of obs.   &  VLBI Array & Freq. [GHz] \\
            \noalign{\smallskip}
            \hline
            \noalign{\smallskip}
  \phantom{0}6 May 1995    & VLBA+Y1$^{\mathrm{a}}$ & 1.6, 5, 15      \\
          26 March 1998    & VLBA+Y1 & 5, 8.4, 15, 22  \\
          18 April 1998    & VLBA+Y1 & 5, 8.4, 15, 22  \\      
	    22 May 1998    & VLBA+Y1 & 5, 8.4, 15, 22  \\
 \phantom{0}6 June 1998    & EVN+MERLIN         & 1.6  \\
                           & +VLBA+Y1           &      \\
           16 June 1998    & VLBA+Y1 & 5, 8.4, 15, 22  \\
           13 Feb. 2000    & EVN+HartRAO+VLBA   & 1.6  \\ 
           20 Feb. 2000    & EVN+HartRAO+VLBA   & 1.6  \\
            27 May 2000    & EVN+HartRAO+VLBA   & 1.6  \\
	    \noalign{\smallskip}
            \hline
         \end{tabular}
\begin{list}{}{}
\item[$^{\mathrm{a}}$] a single element of the VLA
\end{list}
   \end{table}
%__________________________________________________________________

%______________________________________________________________
   \begin{figure*}
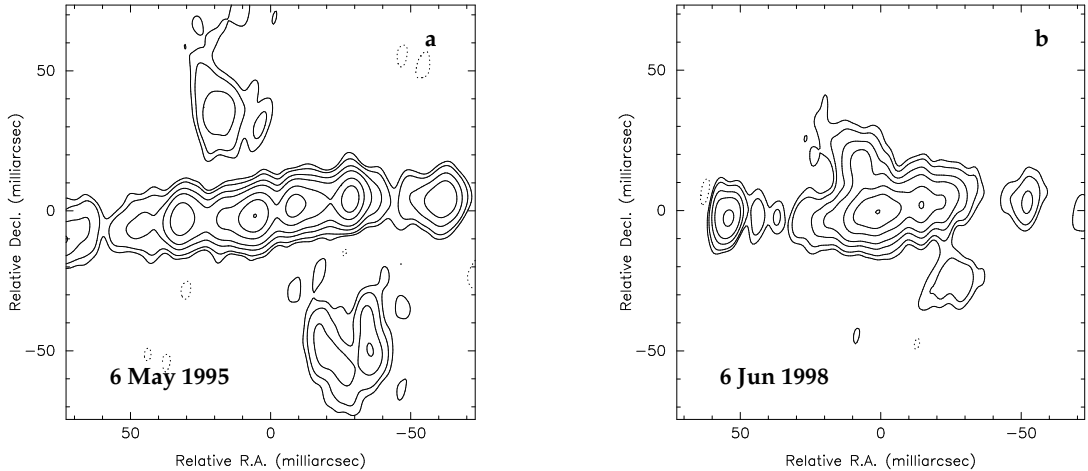

   \centering

   \vspace{20pt}
    \includegraphics[bb=80 100 520 650,clip,angle=-90,width=8cm]{paragi_fig1a.ps}
    \includegraphics[bb=80 100 520 650,clip,angle=-90,width=8cm]{paragi_fig1b.ps}
 
   \caption{The equatorial emission region in 1995 ({\bf a}) and 1998 ({\bf b}) at 1.6 GHz. There are
            changes in the position angle and the separation of the components, but
            the overall structure is similar. The contour levels are at $\pm$1, 2, 5, 10, 
            25, 50, 99\%. The peak flux densities and the restoring beams are {\bf a} 53.6 mJy/beam; 
            11.1$\times$5.6 mas at $PA=-4\deg$, and {\bf b} 38.7 mJy/beam; 10.8$\times$3.8 mas
            at $PA=-5.7 $
            \label{bv5-gp17}
           }
    \end{figure*}
%______________________________________________________________

\section{The Equatorial Emission Region}

The SS433 jets have been studied for two decades. Their kinematic
modell is very well established by optical and radio observations. 
However our data from 1995 observations
showed two emission regions quasi-perpendicular to the beams. The
emission was clearly identified at 1.6 GHz, and even at 5 GHz (apparent
when convolving with the beam of the L-band image or larger). The spectral
index of the emission was remarkably steep ($\alpha\sim-1$). The follow-up
global VLBI experiment in 1998 fully confirmed the existence of 
equatorial radio components with a similar disposition, but at a different
distance from the centre and at a slightly different position angle 
(see Fig.~1). 
The emission spans about 100~mas, corresponding to several hundreds of 
Astronomical Units (1~mas=5~AU at a distance of 5 kpc). A brightness 
temperature of $10^{7}-10^{8}$~K is determined that is indicative of
non-thermal emission (Paragi et al. \cite{IAU205,3rdmicro}).

\section{Indications for outflow}

Observations in other wavebands indicated the presence of an equatorial
outflow in the system. Zwitter et al. (\cite{ZWI91}) invoked a disk-like
outflow to explain the assymetric optical lightcurve, while Kotani et al.
(\cite{KOT96}) suggested a sprinkling disk in order to explain the X-ray
line ratios of Fe\,{\sc xxv} K$\alpha$ red- and blueshifted lines. At the
same time Chakrabarti and Matsuda (\cite{C&M92}) showed by numerical 
simulations that a significant fraction of the accreted matter in SS433
may leave the system through the L$_{2}$ Lagrangian point (for more details
see Paragi et al. \cite{AA99}, and references therein).

Recent work by King et al. (\cite{ARK00}) indicates that SS433 can only avoid 
a common envelope evolutionary phase (i.e. no binary system, no jets)
if most of the accreted matter is expelled by a high radiation pressure.
This scenario invokes a massive equatorial outflow with a mass-loss rate
of about $10^{-4}$~M$_\odot$/yr. Apart from the direct detection in the radio
regime (Paragi et al. \cite{AA99}), there is other recent observational 
evidence supporting this idea. Gies et al. (\cite{DRG02}) studied the UV 
spectrum with the Hubble Space Telescope. They detect H${\alpha}$ and 
He\,{\sc i} stationary emission lines, sometimes showing P-Cygni profile.
These lines originate in an expanding thick disk, embedding the central binary 
system. Fuchs et al. (in press) analyzed IR spectra of the source. 
They also invoke an expanding disk-wind 
scenario, and fit a 150~K free-free component to the spectrum (the IR emission
certainly comes from a larger volume than the UV line emission). The estimated 
outflow rate is in the order of $10^{-4}$~M$_\odot$/yr. These results draw 
our attention to the fact that the mass-loss rate is about 100 times larger
than what we observe in the radio beams!

%______________________________________________________________
   \begin{figure*}
   \centering

   \vspace{0pt}
    \includegraphics[bb=23 272 572 570,clip,width=13cm]{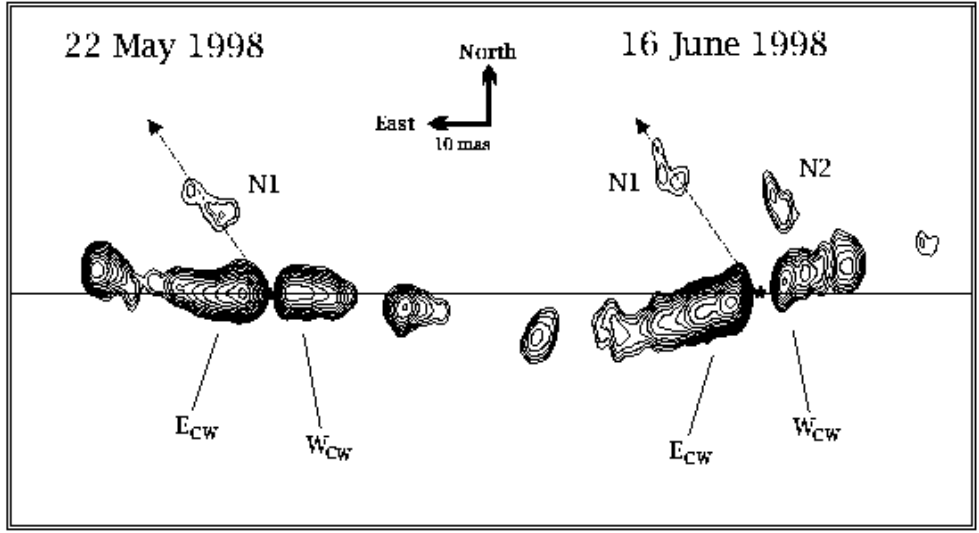}
   \caption{VLBA maps of SS433 at 5 GHz in 1998. The central engine -- indicated
            by an astersik -- is located
            in between the Eastern and Western core-wings (E$_{\rm cw}$ \&
            W$_{\rm cw}$, these correspond to the approaching and the receding
            jet sides at the epochs of observations, respectively). 
            N1 is moving away from the centre with a projected speed of 
            $\sim$1200 km/s. Note that this component is the
            the same as the bright Northern equatorial feature in Fig.~1b
            (from Paragi \cite{PhD01}) 
            \label{pm}
           }
    \end{figure*}
%______________________________________________________________

\section{Direct detection of the outflow?}

In 1998 at two epochs of our multi-frequency VLBA monitoring of SS433, we
identified component N1 at 5~GHz (Fig.~2), at 1.6~GHz detected as a bright
region North to the centre (see Fig.~1b). Our higher resolution at 5~GHz
enables us 
to trace the apparent motion of N1 from the central region. We estimate a
motion of 1200/sin($i$) km/s, where $i$ is the inclination of the
velocity vector of N1 to the line of sight. Although this speed is fairly 
consistent with an early type stellar wind, and similar speeds have been
reported in SS433 (Panferov \& Fabrika \cite{FAB}), we are not yet fully 
convinced that this structural change really represents bulk motion of the 
outflow. Note that we could not find high velocity H\,{\sc i} 
gas in the system with the Westerbork Synthesis Telescope. Note that on 
Fig~2b a new component (N2) appeared at a different position angle.

Observations in 2000 do not allow an accurate proper motion estimate,
but clearly confirm outward motion from the centre (Fig.~3). At the first 
two epochs we see roughly the same structure --- much smoother emission
than in previous experiments. At the third epoch a new component appears,
very close to the centre. So new components always show up close to the
central region. It seems that the appearance of these features
is related to the precessional cycle, and they move away from the central
engine on a timescale of weeks/months.

\section{Nature of emission}

Blundell et al. (\cite{KB01}) recently confirmed the
existence of the equatorial emission. They have observed a smooth structure with
a global VLBI array of VLBA, MERLIN, and the phased VLA at 1.6 and 5 GHz. 
The spectral index of the region at the epoch of observation was flat.
As we have shown above, the appearance of the emission is changing with
time: sometimes it is blob-like, another time it is smoother. An
interesting new result of their experiment is that the spectral index also
changes. Blundell et al. explain the flat spectral index with thermal
free-free emission. Because there is no other observational support for
this idea (e.g. X-rays coming from the region), and because it is difficult to 
explain how the ISM could be heated up to $10^{7}-10^{8}$~K on hundreds of AU 
scales, we consider this scenario to be unlikely. In fact there are observations 
that directly contradict to this model, to mention only one, the observed steep 
spectral index in 1995.

There is a possibility that in fact we observe synchrotron radiation.
Even the flat spectral index can be reconciled with this interpretation, if 
there is a mixed population of relativistic and thermal electrons (see White
(\cite{RLW85}) for details). The question is now how and where these 
energetic electrons are produced. Whether these particles are accelerated
in shocks in the ISM or most of them originate from the vicinity of the
central engine is an open question, but the latter seems to be more
plausible.

% other rjxrbs/most of the mass

%______________________________________________________________
   \begin{figure*}
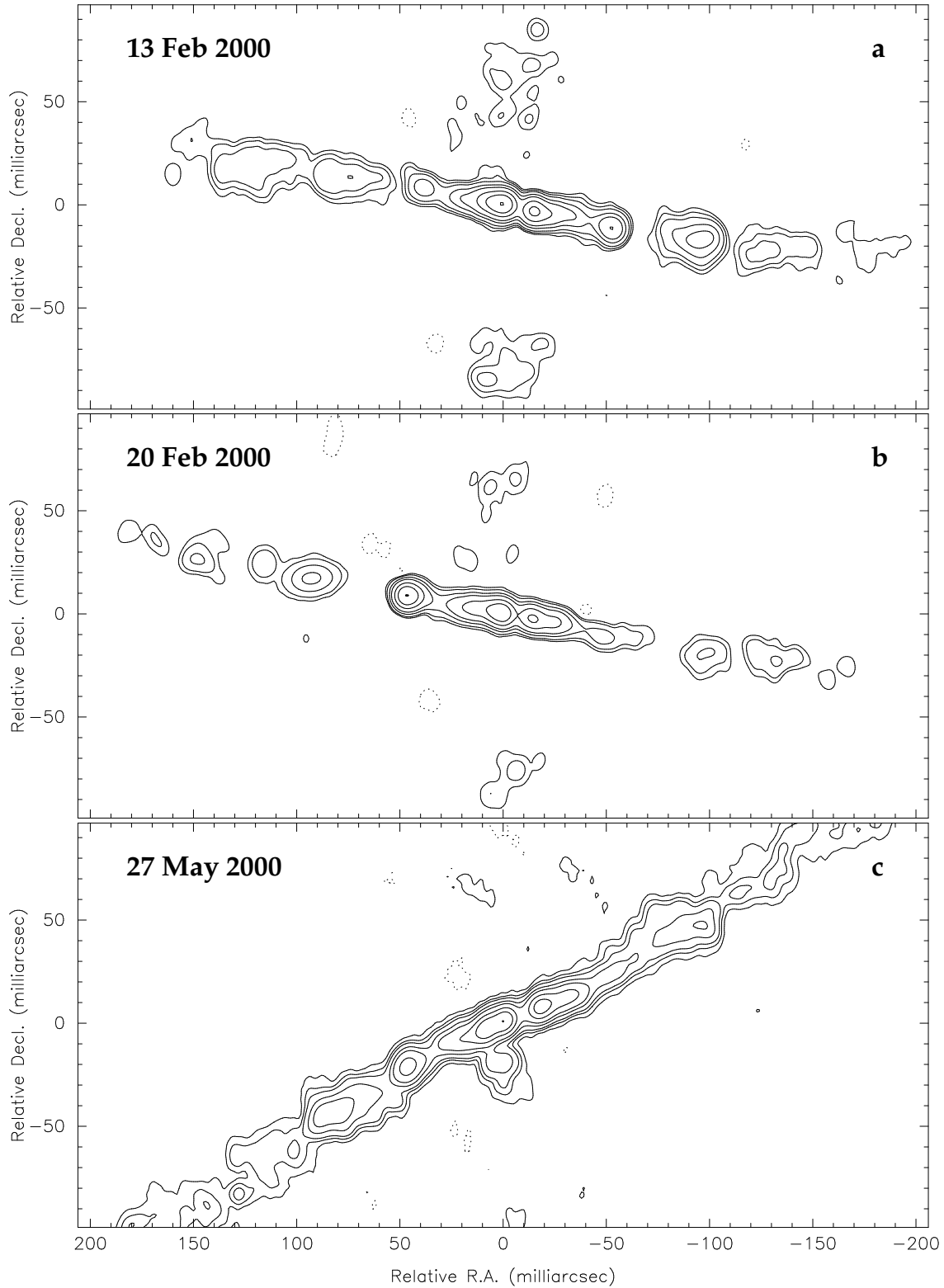

   \centering
   \vspace{20pt}
    \includegraphics[bb=113 34 428 754,clip,angle=-90,width=15cm]{paragi_fig3a.ps}
    \includegraphics[bb=113 34 428 754,clip,angle=-90,width=15cm]{paragi_fig3b.ps}
    \includegraphics[bb=113 34 473 754,clip,angle=-90,width=15cm]{paragi_fig3c.ps}
   \caption{Global VLBI images of SS433 at 1.6 GHz on {\bf a} 13 Feb 2000, {\bf b} 20 Feb 2000,
            and {\bf c} 27 May 2000. The equatorial emission region (to the N and S) 
            is similar at the first two epochs. Note that the Northern and Southern
            emission features are not symmetrical to the brightest component, 
            but to the central engine, located in between the Eastern
            and Western jets. At the third epoch there is a new ``component" emerging from
            the central region. At the full resolution image it is well separated from the
            jet (thanks to Hartebeesthoek). The images were restored with a 8$\times$6~mas
            beam with $PA=0\deg$. The contour levels are $\pm$1, 2, 5, 10, 25, 50, 99\% of
            the peak flux densities of {\bf a} 35.9 mJy/beam (lowest contours are $\pm 0.5$), 
            {\bf b} 27.4 mJy/beam, and {\bf c} 34.1 mJy/beam (lowest contours are $\pm 0.3$)  
            \label{gp025}
           }
    \end{figure*}
%______________________________________________________________

\section{Conclusions}

We detect the equatorial outflow of SS433 that have already been envisaged
by other groups based on optical and X-ray observations, and numerical 
simulations of the system. The outflow speed is estimated to be
1200/sin($i$) km/s, but this result has not been confirmed yet.

Because there are indications for equatorial outflows in other systems 
as well, moreover theoretical work and numerical simulations suggest
that there is a significant mass-loss in high-mass X-ray binary systems,
we conclude that eqatorial outflows might be common in microquasars 
harbouring a massive normal star. In fact, the mass-loss rate could
be much higher in these outflows than in the radio jets, as demonstrated
for SS433.

\begin{acknowledgements}

The European VLBI Network is a joint facility  of European and Chinese 
radio astronomy institutes funded by their national research councils.
The National Radio Astronomy Observatory is operated by Associated Universities, 
Inc. under a Cooperative Agreement with the National Science Foundation.
We acknowledge partial financial support received from the Hungarian Space
Office (MUI), the Netherlands Organization for Scientific Research (NWO), 
and the Hungarian Scientific Research Fund (OTKA) (grant No. N31721 \& T031723). 
This research was supported by the European Commission's TMR Programme
``Access to Large-scale Facilities", under contract No.\ ERBFMGECT950012.
We acknowledge the support of the European Community - Access to Research
Infrastructure and Infrastructure Cooperation Networks (RADIONET, contract No.
HPRI-CT-1999-40003) action of the Improving Human Potential Programme.
\end{acknowledgements}


\begin{thebibliography}{}

\bibitem[2001]{KB01} Blundell K.M., Mioduszewski A.J., Muxlow T.W.B., 
Podsiadlowski P., Rupen M.P. 2001, ApJ 562, L79

\bibitem[1992]{C&M92} Chakrabarti S.M., Matsuda T. 1992, ApJ 390, 639

\bibitem[2000]{ARK00}
King A.R., Taam R.E., Begelman M.C. 2000, ApJ 530, L25

% \bibitem[2002]{FUC02} Fuchs Y., Koch-Miramond L., \ÆAbrah\Æam P. 2002, in the 
% press 

\bibitem[2001]{DRG02} Gies D.R., McSwain M.V., Riddle R.L., Wang, Z., Wiita P.J., 
Wingert D.W. 2002, ApJ 566, 1069

\bibitem[1996]{KOT96} Kotani T., Kawai N., Matsuoka M., Brinkmann W. 1996, PASJ 
48, 619

\bibitem[1989]{M&A89} Margon B., Anderson S.F. 1989, ApJ 347, 448


\bibitem[1997]{FAB} Panferov A.A., Fabrika S.N. 1997, Soviet Astron. 41(4), 506

\bibitem[2000]{Thesis} Paragi Z. 2000, PhD thesis, E\"otv\"os Lor\'and Univ., 
Budapest

\bibitem[2001]{PhD01} Paragi Z. 2001, PADEU 12, p. 53 \\ ({\sc 
http:/$\!$/astro.elte.hu/phd2000/paragi.ps})

\bibitem[2001a]{IAU205} Paragi Z., Fejes I., Vermeulen R.C., Schilizzi R.T., 
Spencer R.E., Stirling A.M. 2001a,
in: Schilizzi R.T., Vogel S., Paresce F., Elvis M. (eds.) Proc. IAU Symposium 
205 ``Galaxies and their constituents
at the highest angular resolutions", p. 266, ASP

\bibitem[2001b]{3rdmicro} Paragi Z., Fejes I., Vermeulen R.C., Schilizzi R.T., 
Spencer R.E., Stirling A.M. 2001b,
Astrophys. Sp. Sc. 276 (suppl.), p. 131, Kluwer Academic Publishers

\bibitem[1999]{AA99} Paragi Z., Vermeulen R.C., Fejes I., Schilizzi R.T., 
Spencer R.E., Stirling A.M. 1999, A\&A, 348, 910

\bibitem[1998]{NewAR98} Paragi Z., Vermeulen R.C., Fejes I., Schilizzi R.T., 
Spencer R.E., Stirling A.M. 1998, NewAR, 42, 641

\bibitem[1985]{RLW85} White R.L. 1985, ApJ 289, 698

\bibitem[1991]{ZWI91} Zwitter T., Calvani M., D'Odorico S. 1991, A\&A 251, 92

\end{thebibliography}
\end{document}